\documentclass[aps,prc,twocolumn,showpacs,superscriptaddress,longbibliography,floatfix,10pt]{revtex4-2}
\usepackage{CJK}
\usepackage{bm}
\usepackage{dcolumn}
\usepackage{amssymb}
\usepackage{amsmath}
\usepackage{graphicx}
\usepackage{dcolumn}
\usepackage{bm}
\usepackage{xcolor}
\usepackage{fix-cm}
\usepackage{mathptmx}
\usepackage[T1]{fontenc}
\usepackage{epstopdf}
\usepackage{subfigure}
\usepackage{booktabs}
\usepackage[colorlinks,citecolor=blue,linkcolor=blue,anchorcolor=blue,filecolor=blue,urlcolor=blue]{hyperref}
\usepackage{multirow}
\usepackage[authormarkup=none]{changes}
\definechangesauthor[name=RA,color=red]{RA}
\makeatletter
\def\NAT@def@citea{\def\@citea{\NAT@separator}}
\makeatother

\begin{document}
\begin{CJK*}{UTF8}{gbsn}
\title{Impact of octupole correlation on the inverse quasifission in  ${}^{160}\text{Gd}+{}^{186}\text{W}$ collisions}
\author{Zhenji Wu ~\CJKfamily{gbsn}(吴振基)}
\affiliation{School of Nuclear Science and Technology, University of Chinese Academy of Sciences, Beijing 100049, China}
\author{Xiang-Xiang Sun ~\CJKfamily{gbsn}(孙向向)}
\affiliation{School of Nuclear Science and Technology, University of Chinese Academy of Sciences, Beijing 100049, China}
\affiliation{Institute for Advanced Simulation (IAS-4), Forschungszentrum J\"{u}lich, D-52425 J\"{u}lich, Germany}
\author{Lu Guo ~\CJKfamily{gbsn}(郭璐)} \email{luguo@ucas.ac.cn}
\affiliation{School of Nuclear Science and Technology, University of Chinese Academy of Sciences, Beijing 100049, China}

\date{\today}

\begin{abstract}	

Multinucleon transfer (MNT) reactions  offer a promising pathway to synthesize neutron-rich heavy nuclei, but the mechanism of inverse quasifission, as a  key reaction channel of MNT, still remains not well understood. We employ time-dependent Hartree-Fock theory to investigate the reaction mechanism, especially the role of the octupole deformed shell in the MNT reaction of ${}^{160}\text{Gd}+{}^{186}\text{W}$.
The results show that inverse quasifission occurs when the deformed projectile and target collide in near tip-tip and tip-side orientations, which favors production of neutron-rich transtarget nuclei.
Interestingly, the distributions and single-particle spectra of primary products reveal that the $N=88$ octupole deformed shell in light fragments dominates inverse quasifission instead of the spherical shells of $^{208}\text{Pb}$ at a center-of-mass energy of $502.6~\text{MeV}$, thus explaining the experimental observation that the yields of the transtarget products are enhanced in the Au region. Further exploration finds that quantum shell effects in inverse quasifission exhibit energy dependence.
These results demonstrate that the octupole deformed shell plays a  crucial role in the inverse quasifission dynamics, significantly advancing the understanding of the MNT reaction mechanism.

\end{abstract}


\maketitle


\section{Introduction}

In a multinucleon transfer (MNT) reaction at energies near the Coulomb barrier, both neutrons and protons can be transferred from the light projectile to the heavy target,
which is termed inverse quasifission~\cite{Zagrebaev2006_PRC73-031602}. 
Unlike quasifission, products in inverse quasifission have larger mass asymmetry than that of the reactants.
Inverse quasifission provides a promising pathway to synthesize neutron-rich heavy and even superheavy nuclei inaccessible through fusion-evaporation reactions and fragmentation reactions~\cite{Adamian2020_EPJA56-47}.
For instance, in the early 1980s, experiments that aimed to probe the production of neutron-rich actinide and transactinide nuclei through inverse quasifission in the ${}^{238}\text{U}+{}^{238}\text{U}$ ~\cite{Gaggeler1980_PRL45-1824} and ${}^{238}\text{U}+{}^{248}\text{Cm}$ ~\cite{Schadel1982_PRL48-852} reactions observed transtarget neutron-rich isotopes of Fm and Md at the level of $0.1~\mu\text{b}$. 
In recent years, the ${}^{238}\text{U}+{}^{232}\text{Th}$ reaction was investigated at Texas A\&M University, and the results suggest that the isotopes with proton numbers as high as 116 might be produced~\cite{Wuenschel2018_PRC97-064602}. 
A new actinide  isotope, $^{241}$U, was identified in the  $^{238}$U+$^{198}$Pt reaction at the KEK Isotope Separation (KISS) facility~\cite{Niwase2023_PRL130-132502}. 
These observations highlight  the prospect of inverse quasifission for producing new neutron-rich heavy and superheavy nuclei.


The reaction mechanism of inverse quasifission has not been well understood due to the complex reaction dynamics involving quantum many-body systems. A prevailing theoretical interpretation proposes that inverse quasifission in actinide reactions is driven by nuclear shell structures, particularly the spherical shells at $Z=82$ and $N=126$ in $^{208}$Pb~\cite{Zagrebaev2006_PRC73-031602}.
Under this interpretation, the large binding energies induced by these spherical shells form valleys in the potential energy surface, driving reaction trajectories toward the valley and enhancing  yields of inverse quasifission products near $^{208}$Pb.
To verify this mechanism, several systems like ${}^{88}\text{Sr}+{}^{176}\text{Yb}$~\cite{Kozulin2014_PRC89-014614} and ${}^{156,160}\text{Gd}+{}^{186}\text{W}$~\cite{Loveland2011_PRC83-44610,Kozulin2017_PRC96-064621} were investigated experimentally. 
These systems were chosen because they are predicted to exhibit inverse quasifission through the same mechanism while offering easier experimental accessibility than actinide systems.
Although the observation of abundant trans-target products confirms the occurrence of inverse quasifission in these reactions, no direct evidence  demonstrates that spherical shells in $^{208}$Pb  drive this process.
Meanwhile, recent studies have increasingly demonstrated that deformed shells play a significant role in nuclear dynamics. 
For instance, octupole deformed shells at $Z=52,~56$~\cite{Scamps2018_Nature564-382,Bernard2023_EPJA59-51} and $N=52,~56,~84,~88$~\cite{Scamps2019_PRC100-041602,Bernard2023_EPJA59-51} have been used to explain nuclear asymmetric fissions.
Deformed shell effects have also been observed in quasifission, where quadrupole deformed shell at $Z=40$~\cite{Umar2016_PRC94-024605} and octupole deformed shells at $Z=52,~84,~88$ and $N=54,~56$~\cite{Godbey2019_PRC100-024610,
McGlynn2023_PRC107-054614} were reported.
While the characteristics of inverse quasifission differ markedly from those of quasifission and fission, investigating whether the deformed shell structures impact inverse quasifission could provide crucial insights into its underlying mechanisms.

Different theoretical approaches have been developed to investigate MNT reactions. Semiclassical methods include the GRAZING model~\cite{Winther1994_NPA572-191}, the complex Wentzel-Kramers-Brillouin (WKB) model~\cite{Szilner2005_PRC71-044610}, the Langevin-type approach~\cite{Zagrebaev2007_JPG34-2265,Zagrebaev2013_PRC87-034608}, and the dinuclear system model~\cite{Adamian2010_PRC81-024604,Wang2012_PRC85-041601,Zhu2022_PLB829-137113,Bao2022_PLB833-137307,Feng2023_PRC108-L051601,Deng2023_PRC107-014616}. Microscopic models comprise the improved quantum molecular dynamics model~\cite{Zhao2009_PRC80-054607,Wen2013_PRL111-012501,Wang2016_PLB760-236,Li2020_PLB809-135697}, the time-dependent Hartree-Fock (TDHF) theory~\cite{Nakatsukasa2016_RMP88-045004,Simenel2018_PPNP103-19,Stevenson2019_PPNP104-142164,Sekizawa2019_FP7-20,Sun2022_CTP74-097302,Simenel2025_EPJA61-181},
and its relativistic extension, time-dependent covariant density functional theory~\cite{Zhang2024_PRC109-024614}.
Recently, the full tensor force has been incorporated in TDHF theory to investigate its role in reaction dynamics~\cite{Dai2014_SCPMA57-1618,Stevenson2016_PRC93-054617,Guo2018_PLB782-401,Guo2018_PRC98-064607,Godbey2019_PRC100-054612,Li2022_PLB833-137349,Sun2022_PRC105-034601,Li2024_PRC110-064607} and fission dynamics~\cite{Huang2024_PRC110-064318}. Among them, TDHF theory provides a self-consistent framework to describe nuclear structure and dynamics based on a unified energy density functional.
It is particularly powerful in elucidating the microscopic mechanisms of nuclear dynamics, such as the influence of quantum shell effects~\cite{Wakhle2014_PRL113-182502,Umar2016_PRC94-024605,Simenel2021_PLB822-136648,Li2022_PLB833-137349,Li2024_PRC110-064607}. 
As such, TDHF has become a crucial tool for investigating various dynamics processes in nuclear physics, including fission~\cite{Simenel2014_PRC89-031601,Goddard2015_PRC92-054610,Tanimura2017_PRL118-152501,Huang2024_PRC110-064318,Huang2024_EPJA60-100}, fusion~\cite{Simenel2004_PRL93-102701,Umar2006_PRC73-054607, Guo2007_PRC76-014601,Guo2008_PRC77-041301,Guo2012_EPJConf38-09003,Washiyama2015_PRC91-064607,Godbey2017_PRC95-011601,Simenel2017_PRC95-031601,Li2019_SCPMA62-122011,Sun2022_PRC105-054610}, quasifission~\cite{Simenel2012_PLB710-607,Yu2017_SCPMA60-092011,Guo2018_PRC98-064609,Li2022_PLB833-137349,Li2024_PRC110-064607,Lee2024_PRC110-024606}, deep inelastic collisions and transfer reactions~\cite{Golabek2009_PRL103-042701,Scamps2013_PRC87-014605,Sekizawa2013_PRC88-14614,Dai2014_PRC90-044609,Wu2019_PRC100-014612,Jiang2020_PRC101-14604,Wu2022_PLB825-136886,Gao2024_PRC109-L041605}.
Using TDHF, it was recently shown that orientation effects of deformed reactants play a role in nucleon transfer of the inverse quasifission reaction ${}^{232}\text{Th}+{}^{250}\text{Cf}$~\cite{Kedziora2010_PRC81-44613}, and deformed shell effects also influence fragment formation for the reaction producing the $^{226}\text{Th}$ compound nucleus~\cite{Lee2024_PRC110-024606}.
Additionally, the stochastic mean-field (SMF) theory~\cite{Ayik2017_PRC96-024611,Sekizawa2020_PRC102-014620} and the time-dependent  random phase approximation theory~\cite{Simenel2011_PRL106-112502,Williams2018_PRL120-022501}, which extend TDHF theory to incorporate quantum fluctuation effects, have been  successfully used to describe the MNT reaction, especially for reactions involving  massive transferred nucleons.

Although significant progress has been made in understanding inverse quasifission, unresolved discrepancies persist between theoretical predictions and experimental observations.
A prominent example is the ${}^{160}\text{Gd}+{}^{186}\text{W}$
 reaction, where the theoretical model, emphasizing the role of the $N=126$ and $Z=82$ shell closures, predicts  enhanced yields of transtarget products near $^{208}$Pb~\cite{Zagrebaev2007_JPG34-2265,Zagrebaev2013_PRC87-034608}. 
However, experimental measurements reveal a striking anomaly: the yield enhancement for trans-target products occurs  in the Au ($Z=79$) region instead of near Pb ($Z=82$)~\cite{Loveland2011_PRC83-44610}.
This inconsistency indicates potential limitations in the current understanding of inverse quasifission dynamics.
To resolve this, further theoretical investigations, particularly at the microscopic level, are essential.
Recent SMF calculations in Ref.~\cite{Ayik2023_PRC108-054605} reproduce the mass distribution of the primary fragments in the ${}^{160}\text{Gd}+{}^{186}\text{W}$ reaction, but their analysis of the reaction mechanism, particularly the influence of shell effects on the inverse quasifission, remains incomplete.
In the present work, we investigate the ${}^{160}\text{Gd}+{}^{186}\text{W}$ reaction using the microscopic TDHF method to elucidate the reaction dynamics, with particular focus on quantum shell effects in inverse quasifission. The paper is organized as follows.
Section \ref{method} shows the TDHF theory, key formulas for primary-product cross sections, and computational details.   
In Sec. \ref{results}, we present the inverse quasifission dynamics and  primary-product distributions for the ${}^{160}\text{Gd}+{}^{186}\text{W}$ reaction. We also discuss the impact of the octupole deformed shell on inverse quasifission.
 Finally, a conclusion of the present work is given in Sec. \ref{conclusion}.

\section{Theoretical framework}\label{method}

In TDHF theory, the  wave function of the many-body system is assumed to be a Slater determinant at all times, which is composed of a series of single-particle wave functions $\phi_k\left(\mathbf{r_i}s_iq_i,t\right)$:
\begin{equation}\label{eq:1}
\Phi(\mathbf{r}_1s_1q_1,\cdots,\mathbf{r}_As_Aq_A,t)=\frac{1}{\sqrt{A!}}\det\left\{\phi_k\left(\mathbf{r}_is_iq_i,t\right)\right\},
\end{equation}
where $A$ is the total number of nucleons and the subscript $k$ labels the single-particle wave functions ($k=1,\ldots,A$). Therefore, the Pauli exclusion is automatically incorporated.
The time evolution of single-particle wave functions  is described by the TDHF equation
\begin{equation}\label{eq:2}
i\hbar\frac{\partial\phi_k\left(\mathbf{r}sq,t\right)}{\partial t} = \hat{h}(t) \phi_k\left(\mathbf{r}sq,t\right),
\end{equation}
which is obtained by employing the time-dependent variation  with respect to the single-particle wave functions. Here, $\hat{h}(t)$ is the single-particle Hamiltonian.
The Skyrme effective interaction is employed to describe the interactions between nucleons.
The parameters of the Skyrme interaction, as the only input parameters of the TDHF simulation, are determined by fitting the properties of static nuclear structure and infinite nuclear matter.
Therefore, TDHF provides a description without free parameters  for  nuclear  dynamic processes, which is crucial for predicting reaction outcomes.
The initial wave function of the reaction system is prepared  by the static Hartree-Fock (HF) method with the same Skyrme effective interaction.

In the TDHF treatment of MNT reactions, the reaction dynamics follows a deterministic and most probable trajectory. Consequently, the proton and neutron numbers of the fragment are the mean values over all possible transfer channels. 
To obtain the transfer probability of each transfer channel, the particle-number projection (PNP) technique~\cite{Simenel2010_PRL105-192701} is used.
The PNP operator is defined as 
 \begin{equation}\label{eq:3}
 \hat{P}_n^V=\frac{1}{2\pi}\int_0^{2\pi}\mathrm{d}\eta e^{i(n-\hat N^V)\eta},
 \end{equation}
 where  $\hat N^V$ is the particle-number operator  in  the subspace $V$. 
For given impact parameter $b$ and center-of-mass energy $E_\text{c.m.}$, the probability $P_{Z,N}$ of finding the fragment with $Z$ protons and $N$ neutrons in the subspace $V$ is given by
 \begin{equation}\label{eq:4}
P_{Z,N}(b,E_\text{c.m.})=\langle\Phi_\text{f}|\hat{P}_N^V|\Phi_\text{f}\rangle\langle\Phi_\text{f}|\hat{P}_Z^V|\Phi_\text{f}\rangle,
 \end{equation}
where $|\Phi_\text{f}\rangle$ is the many-body wave function at the end of  TDHF evolution.
The primary cross section for the fragment with $Z$ protons and $N$ neutrons  is then calculated by integrating the probability $P_{Z,N}$ over the impact parameter $b$, 
\begin{equation}\label{eq:6}
\sigma_{Z,N}(E_\text{c.m.})= 2\pi\int_{b_{\text{low}}}^{b_{\text{up}}}\mathrm{d}bP_{Z,N}(b,E_\text{c.m.})b,
\end{equation}
where  $b_\text{low}$ and $b_\text{up}$ represent the lower and upper bounds of the impact parameter range within which MNT reactions occur.

In this study, the SLy5 Skyrme force~\cite{Chabanat1998_NPA635-231}, incorporating the relevant time-odd and time-even  components in the Hamiltonian, is used for both static and dynamical calculations.
This parameter set has also been widely adopted in recent TDHF studies~\cite{Sekizawa2017_PRC96-014615,Guo2018_PLB782-401,Guo2018_PRC98-064607,Wu2020_SCPMA63-242021,Wu2022_PLB825-136886,Roy2022_PRC105-044611,Sun2022_PRC105-034601,Sun2023_PRC107-L011601,Sun2023_PRC107-064609,Li2024_PRC110-064607,Gao2024_PRC109-L041605}. 
The ground-state wave functions of two colliding nuclei are calculated  on a three-dimensional  $24\times24\times24$ grid with a $1$~fm mesh spacing.
The projectile nucleus $^{160}$Gd exhibits strong quadrupole and slight triaxial and octupole deformations in the ground state, with quadrupole, triaxial, and octupole deformation parameters of $\beta_{2} \approx 0.32$, $\gamma \approx 7.9^\circ$, and $\beta_{3} \approx 0.03$, respectively, while the ground state of target $^{186}$W is prolate deformed with $\beta_{2} \approx 0.24$.
These values agree reasonably well with  experimental data ($\beta_{2} \approx 0.35$ and $ 0.23$)~\cite{Pritychenko2016_ADT107-1}. 
The collision axis is chosen to be parallel to the $x$ axis and the collision occurs in the $x$-$z$ plane. 
The initial separation distance between the projectile and target  is set to be $30$ fm along the colliding axis. 
Prior to TDHF evolution, the colliding nuclei are assumed to move along a Rutherford trajectory  until they reach the initial separation distance.
The TDHF calculations are performed  on a $60\times24\times56$~$\text{fm}^3$ grid without any symmetry restrictions  and the time step  is set to be $0.2$ fm/$c$.
The dynamical evolution is stopped when the distance between the centers of mass of two outgoing fragments  reaches  $34$ fm.

Because both projectile and target nuclei are well deformed, the initial orientations of reactants should be taken into account in the dynamical evolution.
The reactant orientations are characterized by the angles $\theta_\text{P}$  and $\theta_\text{T}$,  which denote the angles between the collision axis and the principal axis of the projectile and target, respectively.
Although the projectile nucleus $^{160}$Gd exhibits slightly triaxial shape, its orientation angle $\theta_\text{P}$ is defined following axial deformed symmetry.
For convenience, the four extreme orientations are denoted as tip-tip ($\theta_\text{P}=0^\circ$, $\theta_\text{T}=0^\circ$), tip-side ($\theta_\text{P}=0^\circ$, $\theta_\text{T}=90^\circ$), side-tip ($\theta_\text{P}=90^\circ$, $\theta_\text{T}=0^\circ$), and side-side ($\theta_\text{P}=90^\circ$, $\theta_\text{T}=90^\circ$). 
In principle, all orientations of the reactants should be considered, but this would be prohibitively time consuming.
To reduce computational costs, we perform TDHF simulations with $\theta_\text{P}$ ranging from $0^\circ$ to $180^\circ$  with increments of $\Delta\theta=22.5^\circ$, covering nine distinct angles.
For the target orientation,  only two angles were considered: $\theta_\text{T} = 0^\circ$ and $90^\circ$.
Thus, a total of 18 distinct initial orientations are simulated in this work.

\section{Results and discussion}\label{results}

\begin{figure}
\includegraphics[width=0.47\textwidth]{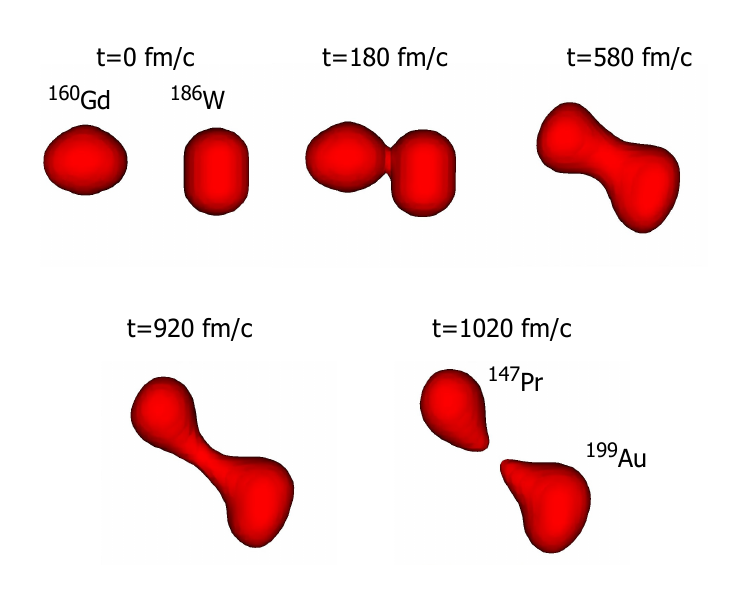}\\
\caption{Time evolution of isodensity surfaces at $0.03~\text{fm}^{-3}$  for a typical inverse quasifission process in the tip-side ($\theta_{\text{P}}=0^\circ$ and $\theta_{\text{T}}=90^\circ$) reaction of ${}^{160}\text{Gd}+{}^{186}\text{W}$  at $E_{\text{c.m.}}=502.6~\text{MeV}$  and $b=2~\text{fm}$. The reaction times, initial reactants, and resulting fragments are labeled.} 
\label{Fig:density}
\end{figure}

In our TDHF simulations, inverse quasifission processes are observed in the $^{160}\text{Gd} +^{186}\text{W}$ reaction.
To provide an overview of inverse quasifission, we show  the density evolution for a typical inverse quasifission event in the tip-side collision at the center-of-mass energy $E_{\text{c.m.}}=502.6$ MeV and impact parameter $b=2$ fm in Fig.~\ref{Fig:density}.
As the projectile nucleus $^{160}$Gd and the target nucleus  $^{186}$W approach each other, they overcome the Coulomb repulsion, converting the kinetic energy into intrinsic excitation.
At $t=180$ fm/$c$, two reactants contact, forming a dinuclear system with a neck. 
Then the neck gradually thickens while the system rotates.
Nucleons are transferred through the neck, accompanied by the dissipation of energy and the transfer of angular momentum.
Over time, the neck elongates due to the combined effects of Coulomb and centrifugal forces. 
Eventually, at $t=1020$ fm/$c$,  the neck breaks and the dinuclear system is split into two fragments: $^{147}$Pr and $^{199}$Au.
About eight neutrons and  five protons are transferred from $^{160}$Gd to $^{186}$W. 

\begin{figure}
\includegraphics[width=0.47\textwidth]{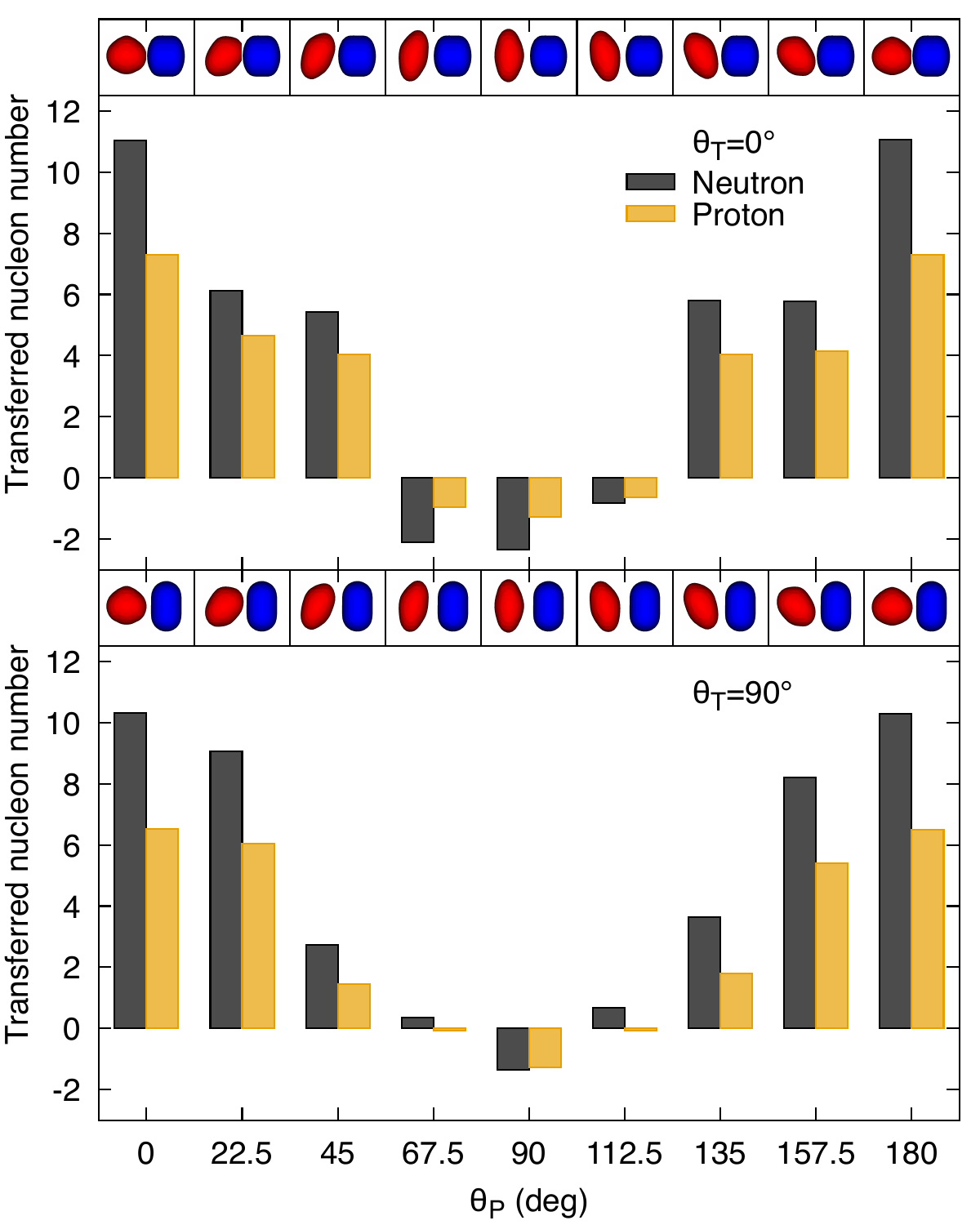}\\
\caption{Transferred nucleon numbers in central collisions of ${}^{160}\text{Gd}+{}^{186}\text{W}$ as a function of initial orientations of projectile at $E_\text{c.m.}=502.6$ MeV. The upper and lower panels are for the tip and side orientations of the target ($\theta_\text{T}=0^\circ$ and $90^\circ$), respectively. The density profiles for various orientations are also shown at the top of each panel.}
\label{Fig:average_nucleon_central}
\end{figure}

The orientations of deformed reactants have a strong  influence on the reaction dynamics~\cite{Hinde1996_PRC53-1290,Zagrebaev2008_IJMPE17-2199,Golabek2009_PRL103-042701,Kedziora2010_PRC81-44613,Wakhle2014_PRL113-182502,Li2022_PLB833-137349,Sun2022_PRC105-034601,Sun2023_PRC107-L011601,Sun2023_PRC107-064609,Li2024_PRC110-064607}.
To investigate the inverse quasifission of  ${}^{160}\text{Gd}+{}^{186}\text{W}$, we show in detail the numbers of transferred protons and neutrons in central collisions with all the above mentioned $18$ initial orientations in Fig.~\ref{Fig:average_nucleon_central}. 
The transferred nucleon number ($\Delta N$) is defined as the difference of nucleon numbers
between the heavy fragments in the final and initial states ($\Delta N=N^\text{H}_\text{final}-N^\text{H}_\text{initial}$).
Therefore, the plus (or minus) sign of the transferred nucleon number indicates nucleon transfer from $^{160}$Gd to $^{186}$W (or from $^{186}$W to $^{160}$Gd). 
For two cases with $\theta_\text{T}=0^\circ$ (upper panel) and $\theta_\text{T}=90^\circ$ (lower panel), the numbers of transferred neutrons and protons exhibit a similar trend with $\theta_\text{P}$ varying from $0^\circ$ to $180^\circ$.
This indicates that the orientations of $^{160}$Gd have a more pronounced impact on the nucleon transfer than those of $^{186}$W.
Inverse quasifission occurs for most orientations, with protons and neutrons transferred from $^{160}\text{Gd}$ to $^{186}\text{W}$, producing transtarget nuclei.
Particularly the number of transferred nucleons decreases as the projectile rotates from tip ($\theta_\text{P}=0^\circ$) to side ($\theta_\text{P}=90^\circ$) orientation, and the slight asymmetry around $90^\circ$ is due to the small octupole deformation of $^{160}\text{Gd}$.
From the case with $\theta_\text{P}=0^\circ$ and $\theta_\text{T}=0^\circ$, the maximum transfer (about $11$ neutrons and $7$ protons) is observed.
Only for collisions around  $\theta_\text{P}=90^{\circ}$, nucleons are transferred from $^{186}$W to $^{160}$Gd along the direction of mass equilibrium. 
In contrast, the charge  equilibration process, in which protons and neutrons typically exhibit opposite transfer directions, is not observed because of the nearly identical neutron-to-proton ratios $N/Z$ of $^{160}$Gd ($N/Z=1.50$) and $^{186}$W ($N/Z=1.51$).

\begin{figure}
\includegraphics[width=0.49\textwidth]{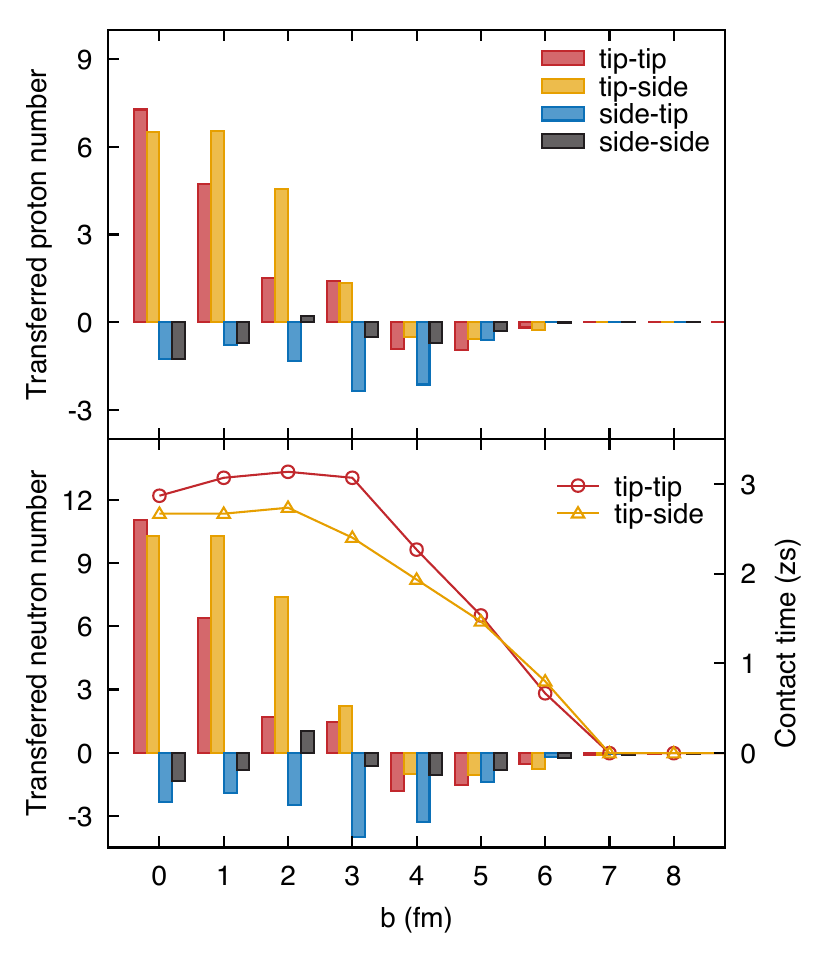}\\
\caption{Transferred proton (upper panel) and neutron (lower panel) numbers as a function of the impact parameter for four extreme initial orientations: tip-tip, tip-side, side-tip, side-side in the reaction ${}^{160}\text{Gd}+{}^{186}\text{W}$ at $E_\text{c.m.}=502.6$ MeV. In the lower panel, contact time (right axis) is displayed for tip-tip (circles) and tip-side (triangles) orientations.}
\label{Fig:average_nucleon_b}
\end{figure}

The nucleon transfer in  noncentral collisions is also analyzed. 
Figure \ref{Fig:average_nucleon_b} shows the transferred proton (upper panel)  and neutron numbers (lower panel) as a function of  the impact parameter at four extreme orientations: tip-tip ($\theta_\text{P}=0^\circ$ and $\theta_\text{T}=0^\circ$), tip-side ($\theta_\text{P}=0^\circ$ and $\theta_\text{T}=90^\circ$), side-tip ($\theta_\text{P}=90^\circ$ and $\theta_\text{T}=0^\circ$), and side-side ($\theta_\text{P}=90^\circ$ and $\theta_\text{T}=90^\circ$).
In the lower panel, the contact time (right axis) is also shown for tip-tip and tip-side orientations.
The Coulomb barriers for these orientations  calculated by the frozen Hartree-Fock method~\cite{Washiyama2008_PRC78-024610,Guo2012_EPJConf38-09003,Guo2018_PLB782-401} are 422.47, 446.07, 460.86, and 486.75 MeV, respectively.
The calculated Coulomb barriers are close to the estimations in Ref. \cite{Kozulin2017_PRC96-064621}. 
Figure \ref{Fig:average_nucleon_b} reveals a clear impact parameter dependence in nucleon transfer.
Inverse quasifission occurs in the tip-tip and tip-side collisions for impact parameters  $b\leq3.5$ fm.
However, when the impact parameter increases to $3.5\leq b\leq7$ fm,  inverse quasifission  disappears and nucleons are  instead transferred from $^{186}$W to $^{160}$Gd, driving the system toward mass equilibration. 
For tip-tip and tip-side orientations, the number of transferred nucleons generally decreases as the impact parameter increases.
This trend is qualitatively associated with the decreasing contact time, although the relationship is not strict, especially at small impact parameters where contact time saturates.
For the side-tip orientation, nucleons are always transferred in the direction of mass equilibration up to $b\leq 7$ fm.
For side-side collisions, only a few nucleons are transferred due to the high Coulomb barrier.
When the impact parameter exceeds $7$ fm in these four cases,  no nucleons are transferred, corresponding to elastic scattering.

Although the experimental data on the primary products of the ${}^{160}\text{Gd}+{}^{186}\text{W}$ reaction are available \cite{Loveland2011_PRC83-44610,Kozulin2017_PRC96-064621}, 
quantitative comparison between  cross sections calculated by the TDHF method and measurements requires careful consideration.
Experimentally, reaction fragments are detected within an angular range of $25^\circ$\textendash$65^{\circ}$ in the laboratory frame.
However, TDHF calculations face challenges in accurately selecting the contributing fragments within this angular range. 
This is because the TDHF theory gives a deterministic outgoing angle for each collision event, rather than a broad distribution.
This limitation stems from the mean-field nature of TDHF theory, which neglects quantum fluctuations and nucleon-nucleon collisions. 
Consequently, the TDHF theory also tends to underestimate product distributions, especially for the case of massive nucleon transfers~\cite{Simenel2018_PPNP103-19,Sekizawa2020_PRC102-014620}. 
Recently for this reaction, the SMF theory, which is based on TDHF and incorporates quantum fluctuations via quantal diffusion approach, has successfully reproduced the mass distributions of primary fragments\cite{Ayik2023_PRC108-054605}; however, the underlying reaction mechanisms for the inverse quasifission, especially the role of the shell closures, are not revealed.

\begin{figure}
\includegraphics[width=0.47\textwidth]{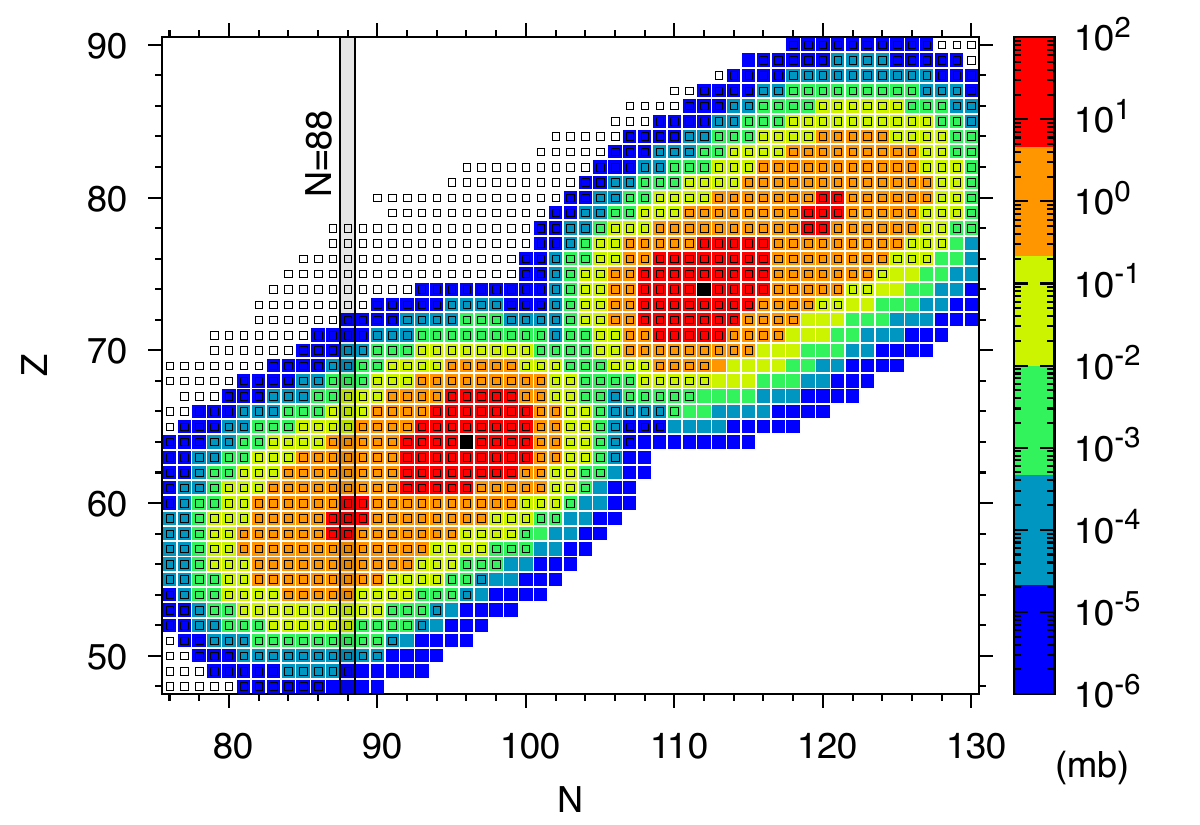}\\
\caption{
Neutron-proton distributions of primary fragments in the tip-side collision of ${}^{160}\text{Gd}+{}^{186}\text{W}$ at $E_\text{c.m.} = 502.6$ MeV.
The projectile and target nuclei are indicated by the black solid squares.
The empty squares represent the nuclei discovered until the year 2021~\cite{Thoennessen_web}. 
The position of the $N=88$ shell is indicated by the gray rectangle.
}
\label{Fig:primary_P0T90}
\end{figure}

The fragment distributions may reveal the dominant driving mechanism of transfer dynamics.
We present in Fig.~\ref{Fig:primary_P0T90} the neutron-proton distribution of primary fragments in the tip-side collision at $E_{\text{c.m.}}=502.6$ MeV by the particle number projection acting on the TDHF wave functions.
This incident energy is same as that used in the experiment~\cite{Kozulin2017_PRC96-064621}.
From Fig.~\ref{Fig:primary_P0T90}, one can see that substantial transtarget nuclei are produced via inverse quasifission processes. 
The obvious enhancements of cross sections induced by inverse quasifission are observed near ${}_{59}^{147}\text{Pr}_{88}$  for light fragments and ${}_{79}^{199}\text{Au}_{120}$ for heavy fragments.
The peak position of heavy fragments is consistent with the experimental observation~\cite{Loveland2011_PRC83-44610} that the yields of
the transtarget products are enhanced in the Au region.
The peak in the enhanced yield of  light fragments occurs at $N=88$, a well-established octupole-deformed shell, indicating that the shell closure at $N=88$ plays the dominant role in inverse quasifission.
However, the $N=126$ and $Z=82$  shell effects in doubly magic nucleus $^{208}\text{Pb}$ exhibit no discernible influence on inverse quasifission at this incident energy.

\begin{figure}
\includegraphics[width=0.45\textwidth]{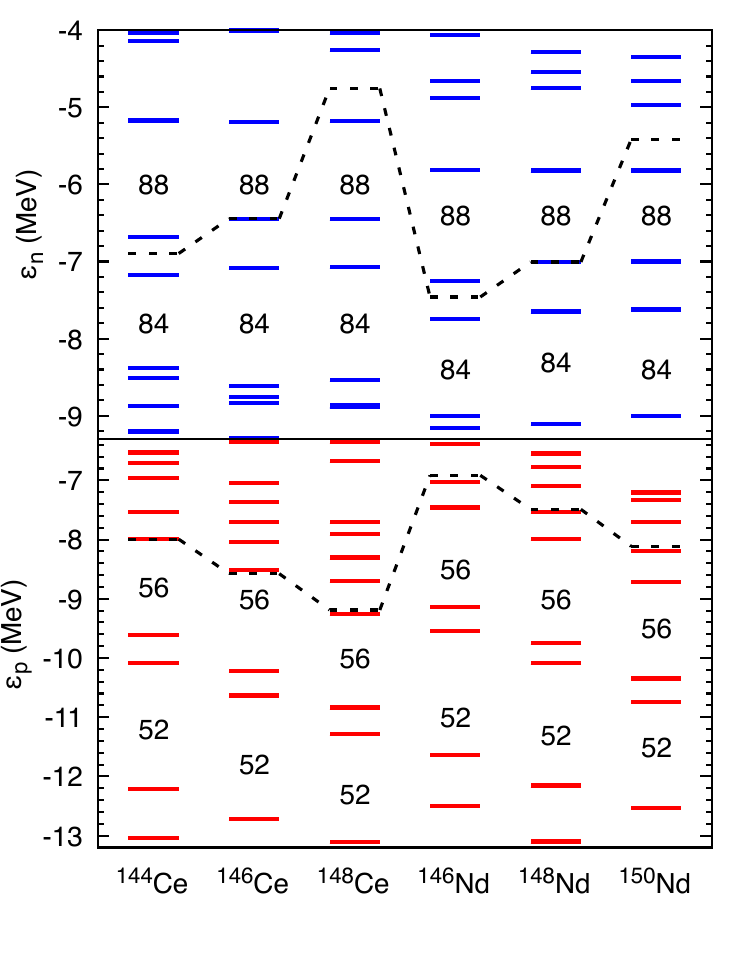}\\
\caption{Single-particle energy levels of the neutron (upper panel) and proton (lower panel) for $^{144,~146,~148}\text{Ce}$ and  $^{146,~148,~150}\text{Nd}$. 
The dashed line denotes the Fermi level. Shell gaps at $N = 84,~88$ and $Z=52,~56$  are labeled.}
\label{Fig:energylevel.}
\end{figure}

The primary fragments are usually strongly deformed at the scission point due to
the large amplitude motion during reaction dynamics (see Fig.~\ref{Fig:density} for a visualization), and naturally the deformed shells of the primary fragments will play a role, which can be reflected by the single-particle structures in the mean-field theories. 
This strategy has been widely used to investigate the deformed shell effects in asymmetric fission~\cite{Scamps2018_Nature564-382,Huang2024_PRC110-064318} and quasifission~~\cite{Li2022_PLB833-137349,Li2024_PRC110-064607}. 
Therefore, in Fig.~\ref{Fig:energylevel.}, we show the neutron 
(upper panel) and proton (bottom panel) single-particle levels 
of even-even nuclei $^{144,146,148}\text{Ce}$ and $^{146,148,150}\text{Nd}$, which are located near the enhancement peaks of light fragments.
Their quadrupole and octupole moments are constrained to be $Q_{20}\simeq3.0~\text{b}$ and $Q_{30}\simeq2.7~\text{b}^{3/2}$, corresponding to the deformation of light fragments at the scission point.
As seen in Fig~\ref{Fig:energylevel.}, pronounced octupole neutron shell gaps at $N=88$ can be found near the neutron fermi energies for all these fragments. These gaps confirm that the $N = 88$ shell closure is the primary driver of inverse quasifission, which leads to the fragment distribution exhibiting a distinct peak centered around $N = 88$ (as observed in Fig.4) for the $^{160}\text{Gd} + ^{186}\text{W}$ reaction at $E_{\text{c.m.}} = 502.6$ MeV.

\begin{figure}
\includegraphics[width=0.47\textwidth]{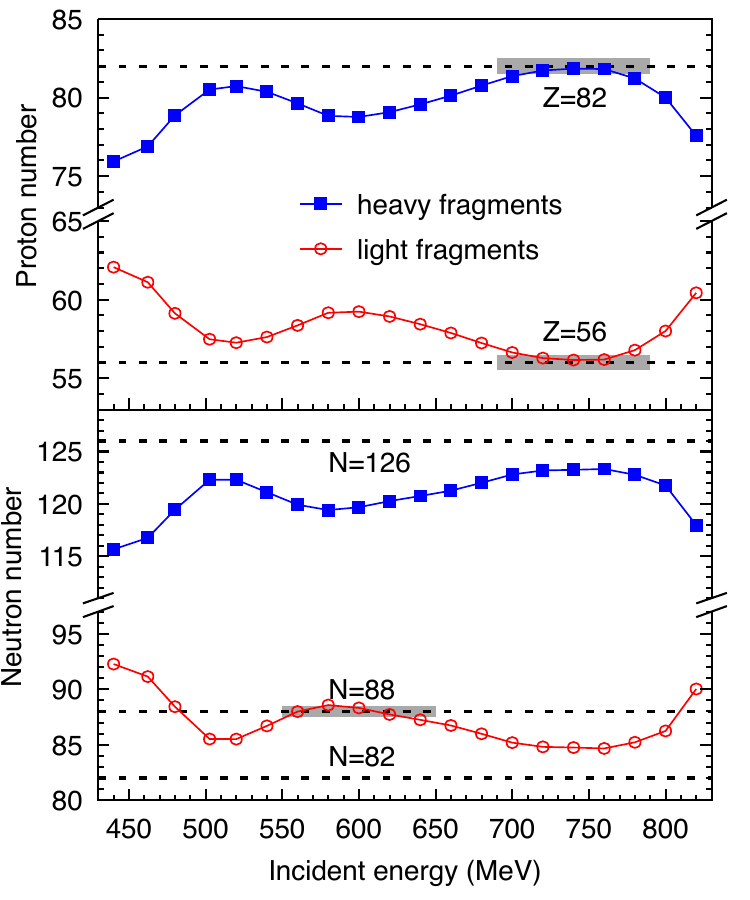}\\
\caption{Average proton (upper panel) and  neutron (lower panel) numbers of light fragments (red open circles) and heavy fragments (blue solid squares)  as a function of incident energy in the central collision of  ${}^{160}\text{Gd}+{}^{186}\text{W}$  at the tip-side orientation. 
The potential shells are indicated by dashed lines.}
\label{Fig:energy.}
\end{figure}

In MNT reactions, when the dinuclear system undergoes scission, the stretched neck causes the resulting fragments to have octupole deformation.
This octupole deformation energetically favors the formation of nuclei with octupole shell correlations, as their deformation energy is minimized in this configuration~\cite{Scamps2018_Nature564-382}.
It is worth mentioning that although $^{208}\text{Pb}$ is spherical in its ground state,  
it readily acquires octupole deformation due to the existence of $3^-$ octupole vibrational states.
That is to say the shells at $^{208}$Pb should principally influence the inverse quasifission in this reaction system.
A natural question arises: Why do the shells of $^{208}\text{Pb}$ not affect the inverse quasifission of ${}^{160}\text{Gd}+{}^{186}\text{W}$ with incident energy 502.6 MeV?
A reasonable explanation is that at this incident energy the system does not have enough energy to transfer more nucleons from Gd to W. Consequently, the heavy fragment products fail to reach the $^{208}\text{Pb}$ region, causing its shell effects to remain inactive in the inverse quasifission process.
To verify this point, we need to study the energy dependence of inverse quasifission of this reaction.

Above, we have already shown that the tip-side orientation is more favorable for nucleon transfer.
Therefore, we calculate the average proton and neutron numbers of fragments in the central collision of $^{160}$Gd + $^{186}$W at center-of-mass energies ranging from 440 to 820 MeV with energy steps of $\Delta E =20~\text{MeV}$, and the results are shown in Fig. \ref{Fig:energy.}.
Inverse quasifission processes happen across the entire energy range, with neutron and proton transfer from $^{160}\text{Gd}$ to $^{186}\text{W}$, and it is clear that the number of transferred nucleons strongly depends on the incident energies.
As the incident energy increases from 440 MeV (near the Coulomb barrier) to 500 MeV, the number of transferred nucleons gradually increases.
In the energy range of $560\text{--}640$ MeV,  the neutron numbers of light fragments mainly concentrate near $N=88$.
When the energy rises further to $700\text{--}780$ MeV, more nucleons are transferred from the projectile to the target nucleus, especially  protons, resulting in heavy fragments toward the spherical $Z=82$ shell, while light fragments naturally populate the octupole deformed $Z=56$ shell region. 
It should be noted that in this energy region 
both the spherical shell at $Z = 82$ in the heavy fragments and the octupole deformed shell at $Z = 56$ in the light fragments may influence the inverse quasifission process. One cannot determine which of spherical or octupole deformed shells or even both are the drivers in the formation of quasifission fragments.
It should be noted that these observations are restricted to central collisions and hence the energy dependence of shell effects mainly reveals the competition between spherical and deformed shells.

\section{CONCLUSIONS}\label{conclusion}
The reaction mechanisms in the ${}^{160}\text{Gd}+{}^{186}\text{W}$ reaction have been studied within the microscopic TDHF approach. 
The energy dependence and the orientation effects of projectile and target are analyzed.
The nucleon transfer is found to be sensitive to  the reactant orientation.  
Collisions with near tip-side and tip-tip orientations preferentially lead to inverse quasifission.
Our calculations first reveal that the $N=88$ octupole-deformed shell in light fragments, rather than the spherical shells of $^{208}$Pb, induces inverse quasifission in the MNT reaction ${}^{160}\text{Gd}+{}^{186}\text{W}$ at the center-of-mass energy $E_\text{c.m.}=502.6~\text{MeV}$. 
This octupole shell effect can explain the experimentally observed yield peak for transtarget products at Au ($Z=79$).
In addition, quantum shell effects in inverse quasifission exhibit energy dependence.
In the energy range of $560\text{--}640~\text{MeV}$, the $N=88$ octupole shell plays a significant role in inverse quasifission, 
while at higher energies of $700\text{--}780~\text{MeV}$, the $Z = 82$ spherical shell competes with the $Z = 56$ octupole deformed shell in the formation of quasifission fragments.

\section*{ACKNOWLEDGMENTS}
We thank Liang Li, Xiangquan Deng, and Yun Huang for helpful discussions. This work has been supported by the National Natural Science Foundation
 of China (Grants No. 12375127, No. 12435008, and No. 12205308), the Strategic Priority Research Program of the Chinese Academy of Sciences (Grant No. XDB1550100), the Fundamental Research Funds for the Central Universities (Grant No. E3E46302), the China Postdoctoral Science Foundation (Grant No. 2024M753173), and the Deutsche Forschungsgemeinschaft (DFG) and NSFC through
the funds provided to the Sino-German Collaborative Research Center TRR110 “Symmetries and the Emergence of
Structure in QCD” (NSFC Grant No. 12070131001, DFG Project-ID 196253076).

\section*{DATA AVAILABILITY}
The data that support the findings of this article are not
publicly available. The data are available from the authors
upon reasonable request.

\end{CJK*}
\bibliographystyle{apsrev4-2}
\bibliography{mybibtexlibrary}

\end{document}